\documentclass[3p,12pt]{elsarticle}
\usepackage{feynmp}
\usepackage{amssymb}
\usepackage{epsfig}
\usepackage{graphicx}
\usepackage{subfigure}
\usepackage{hyperref}
\journal{Nuclear Physics B}

\begin{document}
\begin{frontmatter}
\title{Matter fields near quantum critical point in (2+1)-dimensional U(1) gauge theory}

\author{Guo-Zhu Liu, Wei Li, and Geng Cheng}
\address{\small {\it Department of Modern Physics, University of Science and
Technology of China, Hefei, Anhui, 230026, P.R. China}}

\begin{abstract}
We study chiral phase transition and confinement of matter fields in
(2+1)-dimensional U(1) gauge theory of massless Dirac fermions and
scalar bosons. The vanishing scalar boson mass, $r=0$, defines a
quantum critical point between the Higgs phase and the Coulomb
phase. We consider only the critical point $r=0$ and the Coulomb
phase with $r > 0$. The Dirac fermion acquires a dynamical mass when
its flavor is less than certain critical value $N_{f}^{c}$, which
depends quantitatively on the flavor $N_{b}$ and the scalar boson
mass $r$. When $N_{f} < N_{f}^{c}$, the matter fields carrying
internal gauge charge are all confined if $r \neq 0$ but are
deconfined at the quantum critical point $r = 0$. The system has
distinct low-energy elementary excitations at the critical point
$r=0$ and in the Coulomb phase with $r \neq 0$. We calculate the
specific heat and susceptibility of the system at $r=0$ and $r \neq
0$, which can help to detect the quantum critical point and to judge
whether dynamical fermion mass generation takes place.
\end{abstract}

\begin{keyword}
Gauge theory \sep Chiral symmetry breaking \sep Confinement


\PACS 71.10.Hf \sep 11.30.Qc \sep 11.15.Pg

\end{keyword}

\end{frontmatter}

%%%%%%%%%%%%%%%%%%%%%%%%%%%%%Main Body%%%%%%%%%%%%%%%%%%%%%%%%%%%%%%%%%%%%%

\section{Introduction}
\label{sec:introduction}

Quantum electrodynamics in (2+1)-dimensional space-time has wide
applications in a number of planar strongly correlated systems,
including fractional quantum Hall system, Heisenberg quantum
antiferromagnet, and high-$T_{c}$ cuprate superconductor. Among
its numerous variants, the massless spinor QED$_{3}$ and the
$\mathrm{CP}^{N}$ model are particularly interesting. They both
exhibit fruitful physics in the context of quantum field theory
and have proved to be powerful in understanding some fundamental
physics of cuprate superconductors.

It is widely accepted that the $t$-$J$ model captures the essential
physics of cuprate superconductors \cite{Leereview}. However, it is
far from clear how to treat this model by satisfactory analytic
tools. One frequently used approach to deal with the $t$-$J$ model
is based on the slave-particle technique \cite{Leereview}. For
example, the electron operator on $i$-lattice $c_{i}$ can be
decomposed as
\begin{equation}
c_{i} = f_{i}b_{i}^{\dagger},
\end{equation}
where $f_{i}$ represents the operator of spin-carrying spinon and
$b_{i}^{\dagger}$ represents the charge-carrying holon. Such
decomposition has a redundancy since the electron operator is
invariant under the transformations
\begin{equation}
f_{i}\rightarrow e^{i\theta}f_{i}, \,\,\, b_{i}\rightarrow
e^{i\theta}b_{i},
\end{equation}
which can be considered as an emergent gauge symmetry. This emergent
gauge structure has very important effects on the whole physical
picture and has been investigated extensively for twenty years
\cite{Leereview, Affleck, Ioffe, Kim97, Kim99, Rantner}. However,
such decomposition is not unique and one can alternatively use the
so-called slave-fermion approach, in which the spin quantum number
is carried by scalar field \cite{SenthilSci} and the gauge charge is
carried by fermionic matter field \cite{Kaulprb}.

Recently, a new state of matter, named as algebraic charge liquid,
was proposed to describe the low-energy properties of cuprate
superconductors \cite{Kaulprb, Kaul}. The effective field theory of
this state consists of $N_{f}$-flavor massless Dirac fermions,
$N_{b}$-flavor scalar bosons, and a U(1) gauge field. This effective
theory can be loosely considered as a combination of massless spinor
QED$_{3}$ and $\mathrm{CP}^{N_{b}-1}$ model \cite{Kaul}. It has been
used to understand certain quantum phase transitions and various
exotic properties near the quantum critical points \cite{Kaulprb,
Kaul}. Recently, this model was studied by Kaul and Sachdev
\cite{Kaul} with emphasis put on the anomalous dimensions and some
observable thermodynamic quantities including specific heat and
susceptibility near the quantum critical point between Higgs and
Coulomb phases.

When interacting with U(1) gauge field, there is an interesting
possibility that the massless Dirac fermion might undergo a vacuum
condensation, which then generates a finite mass for Dirac fermion
\cite{Appelquist88, Nash, Dagotto}. This mass breaks the
continuous chiral symmetry possessed by Dirac fermions and leads
to new phenomena. Another characteristic property of QED$_{3}$ is
that the strong gauge field can confine the matter fields that
carry nonzero gauge charges \cite{Burden, Maris95}. In the absence
of scalar field, it is found that the fermion mass can be
generated only for flavor $N_{f} < N_{f}^{c} \approx 3.3$
\cite{Appelquist88}, and confinement takes place once the mass is
generated \cite{Maris95}. If the gauge field also couples to the
scalar field $z$, then the phenomena related to chiral phase
transition and confinement will be modified by the dynamics of
scalar bosons.

In this paper, we study chiral phase transition and confinement of
matter fields in the vicinity of the quantum critical point. The
critical fermion flavor $N_{f}^{c}$ depends on two parameters: the
scalar boson flavor $N_{b}$ and the scalar boson mass $r$. When
$r>0$, the system stays in the Coulomb phase, while $r = 0$ defines
the quantum critical point that separates the Coulomb phase from the
Higgs phase. When the chiral symmetry is dynamically broken, the
massless fermions become massive and are confined if $r \neq 0$;
while the Dirac fermions and $z$ bosons are deconfined at the
quantum critical point $r = 0$. Since the confined and deconfined
matter fields behave quite differently, the Coulomb phase and the
critical point have completely different low-energy elementary
excitations. Thus it is possible to specify the quantum critical
point and to verify whether the chiral symmetry is dynamically
broken by measuring observable physical quantities such as specific
heat and susceptibility. We calculate these quantities within the
effective gauge theory and discuss their behaviors at low
temperature.

The paper is organized as follows. We study the dynamical mass
generation for Dirac fermions in Section 2. The
confinement/deconfinement properties of matter fields is discussed
in Section 3. We then calculate the specific heat and susceptibility
at quantum critical point $r=0$ in Section 4 and in the Coulomb
phase with $r \neq 0$ in Section 5. The last section contains
summary and discussion.

\section{Dynamical fermion mass generation for $r \geq 0$}
\label{sec:dyn_mass}

The action for the interaction between massless Dirac fermion and
U(1) gauge field is
\begin{equation}\label{eq:l_qed3}
\mathcal{S}_{f} = \frac{1}{4}\int d^{3}xF_{\mu\nu}F^{\mu\nu} +
\sum_{i}\int d^{3}x \overline{\psi}_{i}
\gamma^{\mu}\left(\partial_{\mu} - ia_{\mu}\right)\psi_{i},
\end{equation}
where the fermion has $N_{f}$ flavor. The spinor field $\psi$
describes the superconducting quasiparticle that carries physical
electric charge. Here, for generality we keep the kinetic term of
gauge field. Following Kaul and Sachdev \cite{Kaul}, we consider
only \emph{non}-\emph{compact} gauge field and omit the instanton
excitations, so the above field theoretic model might not precisely
match the low-lying theory of cuprate superconductors (there are
theoretical and simulation evidence for the irrelevance of
instantons in the presence of massless Dirac fermions with
sufficiently large flavor $N_{f}$ \cite{Hermele, Ichinose,
Nogueira08}). In order to define chiral symmetry, we adopt a
four-component representation of spinor field $\psi$ and three
$4\times 4$ matrices $\gamma_{\mu}$ ($\mu=0,1,2$) satisfying the
Clifford algebra, $\lbrace \gamma_{\mu},\gamma_{\nu}
\rbrace=2\delta_{\mu \nu}$. Since the fermions are massless, the
action $\mathcal{S}_{f}$ respects the following continuous chiral
symmetry
\begin{equation}
\psi \rightarrow e^{i\varphi \gamma_{3,5}}\psi,
\end{equation}
with $\gamma _{3}$ and $\gamma _{5}$ being two $4 \times 4$ matrices
that anticommute with $\gamma _{\mu}$. The action contains no small
coupling constant and generally can only be treated using the
$1/N_{f}$ expansion.

The free propagator for Dirac fermion is
\begin{equation}
G_{0}^{-1}(p) = i\gamma_{\mu}p^{\mu}.
\end{equation}
The interaction with gauge field modifies it into the complete
fermion Green function
\begin{equation}
G^{-1}(p) = i\gamma_{\mu}p^{\mu} + \Sigma(p),
\end{equation}
where $\Sigma(p)$ is the fermion self-energy function.
Generically, the fermion self-energy can be written as
\begin{equation}
\Sigma(p) = i\gamma_{\mu}p^{\mu}(A(p^{2})-1) + B(p^{2}),
\end{equation}
with $A(p^{2})$ being the wave functional renormalization and
$B(p^{2})$ the fermion mass function. To calculate the fermion
self-energy, we can use either conventional perturbation expansion
method or non-perturbative approach. If the usual perturbation
theory is used, then the self-energy function is proportional to
$\sim \eta \gamma\cdot p\ln(p)$ to the leading order of $1/N_{f}$
expansion \cite{Rantner, Kaul}, which represents only a correction
to the wave function (field operator). When combined with the free
propagator $i\gamma_{\mu}p^{\mu}$, this self-energy modifies the
fermion propagator to a new form
\begin{equation}
G(p) = \frac{i\gamma_{\mu}p^{\mu}}{p^{2-2\eta}},
\end{equation}
which reflects the typical behavior of algebraic charge/spin
liquid \cite{Rantner, Kaul}. This result implies that, to the
leading order, the gauge fluctuation is only a marginal
perturbation and no mass term $B(p^{2})$ is generated. In fact,
such mass term can never be generated to any finite order of
perturbative expansion. This can be understood by the symmetry
arguments. Both the free action (\ref{eq:l_qed3}) and the term
$i\gamma_{\mu}p^{\mu}A(p^{2})$ are chiral symmetric, but the mass
term $B(p^{2})$ is not chiral symmetric. Therefore, it is the
chiral symmetry of the action that prevents the occurrence of mass
term $B(p^{2})$.

\begin{figure}[h]
  \centering
  \subfigure[]{
    \label{fig:pi_z1} %% label for first subfigure
    \includegraphics[width=1.2in]{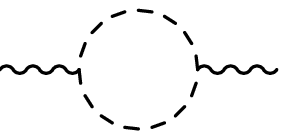}}
      \subfigure[]{
    \label{fig:pi_z2} %% label for first subfigure
    \includegraphics[width=1.0in]{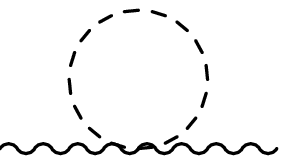}}
      \subfigure[]{
    \label{fig:pi_f} %% label for first subfigure
    \includegraphics[width=1.2in]{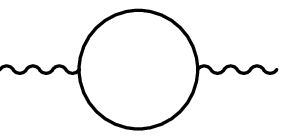}}
      \subfigure[]{
    \label{fig:pi_l} %% label for first subfigure
    \includegraphics[width=1.2in]{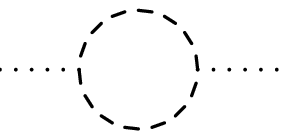}}
\caption{The solid line represents the propagator of Dirac fermions,
dashed line the scalar boson $z$, and dotted line the field
$\lambda$. (a) and (b) correspond to the polarization function from
$z$ bosons. (c) corresponds to the polarization function from Dirac
fermions. (d) corresponds to the polarization of field $\lambda$.}
\label{fig:pi} %% label for entire figure
\end{figure}

However, the perturbation theory can not tell us everything.
Although the action is chiral symmetric, it is possible that its
ground state does not respect the chiral symmetry and the Dirac
fermion acquires a finite dynamical mass. Indeed, the generation of
nonzero $B(p^{2})$ signals the happening of a phase transition and
the reconstruction of the ground state, which are surely
non-perturbative in nature and can not be analyzed by ordinary
perturbation theory. The standard non-perturbative approach to the
problem of dynamical fermion mass generation is to solve the
Dyson-Schwinger (DS) equation for the fermion mass function, which
was originally proposed by Nambu and Jona-Lasinio in the context of
particle physics \cite{Nambu}. The DS equation can be written in the
form
\begin{eqnarray}
G^{-1}(p) = G_{0}^{-1}(p) - \int\frac{d^{3}k}{(2\pi)^{3}}\gamma
_{\mu}G(k)\Gamma_{\nu}(p,k)D_{\mu\nu}(p-k), \nonumber
\end{eqnarray}
where $\Gamma _{\nu }(p,k)$ is the vertex function and
$D_{\mu\nu}(p-k)$ the photon propagator. After a series of
manipulations, the above equation is decomposed into a couple of
equations for $A(p^{2})$ and $B(p^{2})$ respectively
\begin{eqnarray}
A(p^{2}) &=& 1+\frac{1}{4p^{2}}\int
\frac{d^{3}k}{(2\pi)^{3}}\mathrm{Tr}\lbrack \gamma\cdot
k\gamma_{\mu}G(k)\Gamma_{\nu}D_{\mu\nu}(p-k)\rbrack, \nonumber \\
B(p^{2}) &=& -\frac{1}{4}\int
\frac{d^{3}k}{(2\pi)^{3}}\mathrm{Tr}\lbrack \gamma_{\mu}G(k)\Gamma
_{\nu}D_{\mu\nu}(p-k)\rbrack. \nonumber
\end{eqnarray}
If the DS equation for $B(p^{2})$ has only vanishing solution, the
fermions remain massless and are stable against gauge
fluctuations. Whenever the equation develops a nontrivial solution
$B(p^{2})$, the massless fermion acquires a finite mass. The
crucial difference between the perturbative and non-perturbative
approaches is that the DS equation is nonlinear which makes it
possible for chiral phase transition to happen at the bifurcation
point.

The logic presented above is analogous to the gap generation for
fermion excitations in the BCS theory of conventional
superconductors. In a normal metal, when we compute the mean value
$\langle c_{k\uparrow}c_{-k\downarrow}\rangle$ using ordinary
perturbative expansion, only vanishing result can be obtained. It
could be understood that the local gauge symmetry of the system
prevents the appearance of nonzero $\langle
c_{k\uparrow}c_{-k\downarrow}\rangle$ which is not gauge invariant.
However, a finite gap can be obtained by solving the nonlinear gap
equation that is constructed by non-perturbative approach.
Apparently, the local gauge symmetry does not prevent the gap
generation; on the contrary, this symmetry is spontaneously broken
by the finite gap since $\langle c_{k\uparrow}c_{-k\downarrow}
\rangle \neq 0$. The gap generation signals a phase transition from
normal metal to superconductor. In fact, the idea of dynamical
fermion mass generation was motivated by the Cooper pairing picture
in BCS theory \cite{Nambu}. The order parameter for chiral phase
transition is just the fermion mass term
$\langle\bar{\psi}\psi\rangle$ which is caused by fermion vacuum
condensation.

For the spinor QED model (\ref{eq:l_qed3}), the chiral phase
transition has been studied for many years. It is known that there
exists a critical fermion flavor $N_{f}^{c}$ which separates the
chiral symmetric phase ($N_{f} > N_{f}^{c}$) from the chiral
symmetry broken ($N_{f} < N_{f}^{c}$) phase \cite{Appelquist88,
Nash, Dagotto}. The critical flavor is about $N_{f}^{c} \approx 3.3$
to the leading order of $1/N_{f}$ expansion \cite{Appelquist88}.

We now add the action of complex scalar field. At half-filling,
the cuprate superconductor is a Mott insulator with long-range
Neel order. The Neel order parameter $\mathbf{n}$ can be
represented by the scalar boson operator $z_{\alpha}$ by
\begin{equation}
\mathbf{n} = z_{\alpha}^{\ast}\mathbf{\sigma}_{\alpha\beta}
z_{\beta}.
\end{equation}
Then the Neel state can be described by an effective field theory
called $\mathrm{CP}^{N_{b}-1}$ model, where $N_{b}$ is the flavor of
scalar field $z_{\alpha}$. This model admits an emergent local gauge
symmetry, as demonstrated by Polyakov \cite{Polyakov}, and has been
successfully used to describe the planar Heisenberg quantum
antiferromagnetism \cite{SenthilSci, Kaulprb, Kaul}. Its action is
\cite{SenthilSci, Kaulprb, Kaul}
\begin{equation}
\mathcal{S}_{b} = \frac{1}{g}\int d^{3}x \left[|(\partial_{\mu
}-ia_{\mu})z_{\alpha}|^{2} - i\lambda(|z_{\alpha}|^{2}-1)\right].
\end{equation}
The Lagrangian multiplier $\lambda$ is introduced to impose the
constraint $|z_{\alpha}|^{2}=1$. It takes on a uniform saddle point
value that extremizes the action $\lambda_{0} = ir$,
\begin{equation}
\int \frac{d^{3}p}{(2\pi)^{3}}\frac{1}{p^{2}+r} = \frac{1}{g}.
\end{equation}
For $g > g_{c}$, the system stays in the Coulomb phase with
$\langle z_{\alpha}\rangle = 0$; for $g < g_{c}$, the system is in
the Neel phase where the SU(N$_{b}$) rotational symmetry is broken
by $\langle z_{\alpha}\rangle \neq 0$. In this paper, we only
consider the phase with $\langle z_{\alpha}\rangle = 0$. At the
$N_{b}\rightarrow \infty$ limit, the quantum critical point
locates at $r=0$, where $g = g_{c}$ satisfying
\begin{equation}
\int \frac{d^{3}p}{(2\pi)^{3}}\frac{1}{p^{2}} = \frac{1}{g_{c}}.
\end{equation}
From these formula, we have
\begin{eqnarray}
\frac{1}{g_{c}} - \frac{1}{g} = \int
\frac{d^{3}p}{(2\pi)^{3}}\left(\frac{1}{p^{2}}-\frac{1}{p^{2}+r}\right)
= \int \frac{d^{3}p}{(2\pi)^{3}}\frac{r}{p^{2}(p^{2}+r)} =
\frac{\sqrt{r}}{4\pi},
\end{eqnarray}
from which the mass parameter
\begin{eqnarray}
r = \frac{\Lambda^{2}}{\pi^{2}}\frac{(g-g_{c})^{2}}{g^{2}}.
\end{eqnarray}
For finite $N_{b}$, the scalar field $z$ receives self-energy
correction from gauge fluctuations, so the critical point is no
longer at $r=0$. Even in this case, one can define a new quantity
$r_{g}$ so that the system goes critical as it vanishes \cite{Kaul}.
Here, we simply use the single parameter $r$ to denote the boson
mass.

The additional coupling between gauge field and scalar field will
affect the critical flavor $N_{f}^{c}$. The DS equation of fermion
self-energy function will be studied within the $1/N_{f}$ expansion.
In the Landau gauge, the gauge boson propagator has the expression
\begin{equation}
D_{\mu\nu}(q) = \frac{1}{q^{2}[1+\Pi_{a}(q^{2})]}
\Big(\delta_{\mu\nu} - \frac{q_{\mu}q_{\nu}}{q^{2}}\Big),
\end{equation}
where the polarization function is given by
\begin{equation}
\Pi_{a}(q^{2}) = \Pi_{a}^{f}(q^{2}) + \Pi_{a}^{b}(q^{2}),
\end{equation}
with the fermion contribution $\Pi_{a}^{f}(q^{2}) = N_{f}/8|q|$
(Fig. \ref{fig:pi_f}). The scalar sector will be studied using the $1/N_{b}$
expansion. In particular, the scalar boson contribution to one-loop
vacuum polarization (Fig. \ref{fig:pi_z1} and Fig. \ref{fig:pi_z2}) is
\begin{eqnarray}
\Pi_{a}^{b}(q^{2}) = \frac{N_{b}}{4\pi}\Big(\frac{q^{2} +
4r}{2q}\arctan \frac{q}{2\sqrt{r}} - \sqrt{r} \Big).
\end{eqnarray}
Inserting the propagator $D_{\mu\nu}(q)$ into the mass equation,
then the critical fermion flavor $N_{f}^{c}$ becomes a function of
$r$ and $N_{b}$. We now determine this dependence by solving the
mass equation.

To the lowest-order, the wave function renormalization can be simply
taken to be $A(p^{2})=1$ since the next-leading order is suppressed
by a factor of $1/N_{f}$. In the Landau gauge, the vertex function
reduces to $\Gamma_{\mu}(p,k) = \gamma_{\mu}$, as required by the
Ward identity. Under these approximations, the mass equation is
relatively simple. To go beyond this approximation, one should solve
the coupled equations for $A(p^{2})$ and $B(p^{2})$. Further, once
$A(p^{2})\neq 1$, the vertex function can no longer be taken to be
$\gamma_{\mu}$. It must be chosen properly to be consistent with the
Ward identity. As it turns out, the DS equations become much more
complicated after including these higher order corrections. It is
also difficult to chose a proper vertex function. Fortunately, there
is a very useful technique \cite{Landau, Georgi} which simplifies
the calculation. To use this technique, we first write the following
propagator for gauge boson
\begin{equation}
D_{\mu\nu}(q) = \frac{1}{q^{2}[1+\Pi_{a}(q^{2})]}
\Big(\delta_{\mu\nu} - \xi(q^{2})
\frac{q_{\mu}q_{\nu}}{q^{2}}\Big),
\end{equation}
where $\xi(q^{2})$ is a gauge fixing parameter that depends on
momentum. Taking advantage of the gauge symmetry, one can choose a
so-called non-local gauge $\xi(q^{2})$ so that the wave function
renormalization is strictly unity, $A(p^{2}) \equiv 1$. Then the
vertex function becomes $\Gamma_{\mu}(p,k) = \gamma_{\mu}$. The
gauge parameter is determined by the following expression
\begin{equation}
\xi(q^{2}) = \frac{2(1+\Pi_{a}(q^{2}))}{q^{2}}\int_{0}^{q^{2}}
\frac{dt}{1+\Pi_{a}(t)} - 1.
\end{equation}
Now we only need to solve a single DS equation for mass function
\begin{eqnarray}
B(p^{2})=\int\frac{d^{3}k}{(2\pi)^{3}}\frac{B(k^{2})}{k^{2} +
B^{2}(k^{2})}\frac{3-\xi((p-k)^{2})}{(p-k)^{2}(1+\Pi_{a}((p-k)^{2}))}.
\end{eqnarray}
This integral equation is much more easily solved numerically than
the coupled equations of $A(p^{2})$ and $B(p^{2})$. It reduces to
the mass equation in Landau gauge when the gauge parameter $\xi =
1$. There must be a critical fermion flavor $N_{f}^{c}$ below
which a finite fermion mass is generated.

\begin{figure}[h]
  \centering
  \subfigure[]{
    \label{fig:landau} %% label for first subfigure
    \includegraphics[width=3.0in]{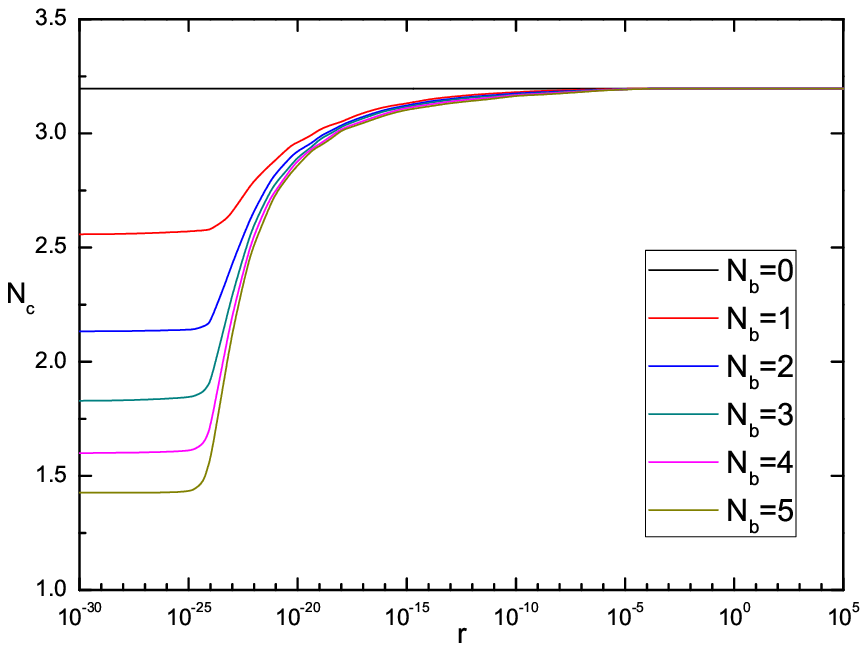}}
      \subfigure[]{
    \label{fig:nonlocal} %% label for first subfigure
    \includegraphics[width=3.0in]{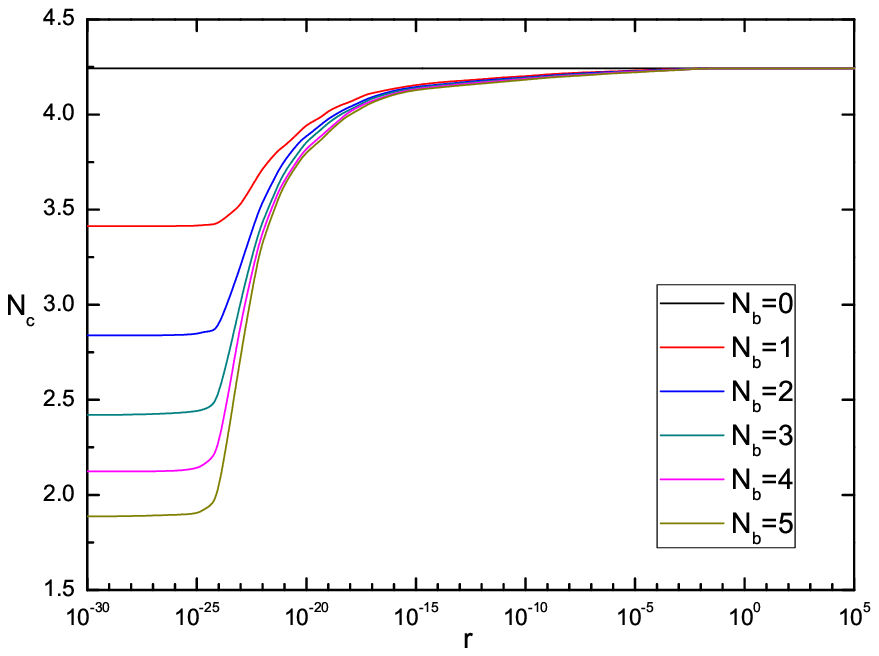}}
\caption{(a) Dependence of critical flavor $N_{f}^{c}$ on scalar
boson mass $r$ for different values of boson flavor $N_{b}$ in the
Landau gauge. (b) Dependence of critical flavor $N_{f}^{c}$ on
scalar boson mass $r$ for different values of boson flavor $N_{b}$
in the non-local gauge.}
\label{fig:nc} %% label for entire figure
\end{figure}

We solved the DS equation \cite{Liu03} in both the Landau gauge and
the non-local gauge, with results presented in Fig. \ref{fig:landau}
and Fig. \ref{fig:nonlocal} respectively. Note that the parameter $r$
now becomes dimensionless since it is re-scaled by some intrinsic
energy scale (the parameter $\alpha$ defined in the paper of
Appelquist \emph{et} \emph{al.} \cite{Appelquist88}). The results
show that the boson mass $r$ increases the critical fermion flavor
$N_{f}^{c}$ and that the boson flavor $N_{b}$ reduces $N_{f}^{c}$
for small mass $r$. When $r$ is of the order unity, the effect of
scalar boson $z_{\alpha}$ on $N_{f}^{c}$ drops out and the critical
flavor $N_{f}^{c}$ is independent of $r$ and $N_{b}$. This is easy
to understand because for large $r$ the scalar boson contribution to
polarization $\Pi_{a}^{b}\rightarrow N_{b}/r$ can be neglected
compared with the fermion contribution $\Pi_{a}^{f}(q^{2}) =
N_{f}/8|q|$ at low momenta. If we believe that the results obtained
in the non-local gauge is more reliable than that in Landau gauge,
then the critical flavor $N_{f}^{c} > 2$ for $N_{b}\leq 4$. When the
physical fermion flavor $N_{f}=2$, a finite fermion mass is always
generated for $N_{b}\leq 4$.

Once the fermion mass is generated, the system is no longer critical
even at the quantum critical point $r=0$. The nonzero value of
$\langle\bar{\psi}\psi\rangle$ corresponds to the formation of
charge density wave since the fermion field $\psi$ carries only
electric charge. As the result of dynamical breaking of continuous
chiral symmetry, there appear massless Goldstone bosons which are
fermion-antifermion pairs. The finite fermion mass has substantial
effects on various observable physical quantities, to which we turn
in the next three sections.

When the scalar bosons condense, $\langle z_{\alpha}\rangle \neq 0$,
the gauge boson becomes massive due to Anderson-Higgs mechanism. The
effect of gauge boson mass on $N_{f}^{c}$ has been discussed
previously in Ref. \cite{Liu03}, where it was found that this mass
rapidly suppresses the dynamical mass generation for Dirac fermions.
We will not repeat such considerations in the present paper.

When applying the effective field theory to realistic cuprate
superconductors, the Dirac fermions actually have two velocities,
$v_{F}$ and $v_{\Delta}$, with $v_{F}/v_{\Delta} \approx 20$
\cite{Leereview}. Although the velocity anisotropy is important in
certain physical quantities, it was shown \cite{Franz} to be
irrelevant near the critical point of chiral phase transition.
Indeed, in most publications on spinor QED$_{3}$ the velocities are
simply set to unity, i.e., $v_{F} = v_{\Delta} = 1$ \cite{Leereview,
Kim97, Kim99, Rantner, Kaul}.

\section{Classical potential of gauge field: confinement and deconfinement}
\label{sec:cl_potential}

The matter fields interact with each other through the mediation
of U(1) gauge field. The classical potential between two particles
with opposite gauge charges can be written in the coordinate space
as \cite{Burden}
\begin{equation}
V(\mathbf{x}) = -\int\frac{d^{2}q}{(2\pi)^{2}}
e^{i\mathbf{q}\cdot\mathbf{x}}\frac{1}{\mathbf{q}^{2}[1+\Pi(\mathbf{q}^{2})]}.
\end{equation}
It has the following asymptotic form \cite{Burden}
\begin{equation}
V(\mathbf{x}) \sim
\frac{1}{2\pi}\frac{1}{1+\Pi(0)}\ln(|\mathbf{x}|).
\end{equation}
Obviously, the potential is determined by the infrared behavior of
vacuum polarization function. In the absence of vacuum polarization
$\Pi(q^{2})$, the potential has a logarithmical form
\begin{equation}
V(\mathbf{x}) \sim \ln(|\mathbf{x}|).
\end{equation}
This potential increases at large distances and hence is a confining
potential, though weaker than a linear potential. When the
polarization $\Pi(q^{2})$ is included, the gauge potential will be
changed.

In cases without scalar field, $N_{b}=0$, the form of gauge
potential $V(\mathbf{x})$ depends only on whether the fermion mass
is zero or finite \cite{Maris95}. For massless fermion, the vacuum
polarization is $\Pi(q^{2}) = \frac{N_{f}}{8|q|}$. Since
$\Pi(0)\rightarrow \infty$, the confining potential is destroyed.
After the fermion mass is generated, the polarization function
$\Pi_{a}^{f}(q^{2})$ becomes
\begin{eqnarray}
\Pi_{a}^{f}(q^{2}) = \frac{N_{f}}{4\pi q^{2}}\Big(2m+\frac{q^{2} -
4m^{2}}{q}\arcsin \frac{q}{\sqrt{q^{2}+4m^{2}}}\Big). \nonumber
\end{eqnarray}
In the limit $q \rightarrow 0$, it has a finite value
\begin{eqnarray}
\Pi_{a}^{f}(0) = \frac{N_{f}}{8\pi m}.
\end{eqnarray}
Thus in the symmetry breaking phase, the gauge potential is
confining and there can not be any asymptotic fermion states
\cite{Maris95}.

Note the gauge field considered here is \emph{non}-\emph{compact}
and there are no topological instanton configurations. The
confinement is induced completely by the gauge force between
particles with opposite charges. It is a special feature of gauge
field in two spatial dimensions.

However, the gauge potential $V(\mathbf{x})$ is modified by the
additional scalar boson. Depending on the fermion mass $m$ and the
scalar boson mass $r$, there are four possibilities:

%\begin{description}
%\item[1]    if $m=r=0$, $\Pi(q^{2})=(2N_{f}+N_{b})/16|q|$ which diverges at $p\rightarrow 0$;
%\item[2]    if $m\neq 0$ and $r = 0$, $\Pi(0) = N_{f}/m+N_{b}/16|q|$ which diverges at $q \rightarrow 0$;
%\item[3]    if $m=0$ and $r\neq 0$, $\Pi(0) = N_{f}/8|q| + N_{b}/r$ which diverges at $p \rightarrow 0$;
%\item[4]    if $m\neq 0$ and $r\neq 0$, $\Pi(0) = N_{f}/m + N_{b}/r$ which is finite.
%\end{description}

\begin{enumerate}
\item if
$m=r=0$, $\Pi(q^{2}) = \frac{2N_{f}+N_{b}}{16|q|}$ diverges at $q
\rightarrow 0$;
\item if $m\neq 0$ and $r = 0$, $\Pi(0) =
\frac{N_{f}}{m} + \frac{N_{b}}{16|q|}$ diverges at $q \rightarrow
0$;
\item if
$m=0$ and $r\neq 0$, $\Pi(0) = \frac{N_{f}}{8|q|} + \frac{N_{b}}{r}$
diverges at $q \rightarrow 0$;
\item if $m\neq 0$ and $r\neq 0$,
$\Pi(0) = \frac{N_{f}}{m} + \frac{N_{b}}{r}$ is finite.
\end{enumerate}

Compared with the case without scalar bosons, there appears a new
possibility: at exactly the critical point where $r=0$, the gauge
potential is always deconfined, irrespective of whether the chiral
symmetry is broken or not. This implies that, in the chiral symmetry
broken phase, the matter fields carrying internal gauge charges are
deconfined at the quantum critical point $r=0$, but are always
confined once the system enters the Coulomb phase with $r \neq 0$.
The deconfined matter fields will make finite contribution to
observable physical quantities, such as specific heat and
susceptibility, but the confined matter fields make no contributions
to them. In the confining phase, only excitations with zero gauge
charge can be mobile and make finite contributions to observable
quantities. Obviously, the quantum critical point and the Coulomb
phase have rather different low-energy elementary excitations. It is
possible to distinguish the critical point from the Coulomb phase by
measuring the observable quantities. These quantities can also be
used to judge whether dynamical chiral symmetry breaking happens or
not.

\section{Deconfinement and observable quantities at the quantum critical point $r=0$}
\label{sec:Deconfinement}

We first discuss the quantum critical point, $r=0$, where the Dirac
fermions and $z$ bosons are deconfined. In this point, the
low-energy elementary excitations are: massive Dirac fermion $\psi$,
massless scalar boson $z$, and composite massless Goldstone boson.
The gauge field is strongly interacting with Dirac fermions and
scalar bosons, so it makes finite contributions to the specific heat
and susceptibility.

We first calculate the free energy $\mathcal{F}$ of the system,
which is given by
\begin{eqnarray}
\mathcal{F} = N_{b}f^{0b} + N_{f}f^{0f} + f^{G},
\end{eqnarray}
where $f^{0b}$, $f^{0f}$, and $f^{G}$ are the free energy of free
scalar bosons, free fermions, and Goldstone bosons, respectively.
To calculate the susceptibility, we follow the strategy of Ref.
\cite{Kaul} and introduce two magnetic fields, $H_{b}$ and
$H_{f}$. For $z$ bosons, the field shifts frequency as
$\omega_n\rightarrow \omega_n - \theta H_b/2$, where $\theta=\pm
1$. For fermions, the coupling is the same, with the replacement
$\omega_{n} \rightarrow \omega_{n}-\theta H_{f}$. Then the
susceptibility $\chi$ and specific heat $C_{V}$ can be computed by
\begin{eqnarray}
\chi_{b} &=& \frac{\partial^{2}\mathcal{F}}{\partial H_{b}^{2}},
\\
\chi_{f} &=& \frac{\partial^{2}\mathcal{F}}{\partial H_{f}^{2}},
\\
C_{V} &=& -T \frac{\partial^{2}\mathcal{F}}{\partial T^{2}}.
\end{eqnarray}
If the system is completely critical, the specific heat has the
typical behavior $\sim T^{2}$. Once the Dirac fermion gets massive,
the system is surely no longer critical and the $\sim T^{2}$
behavior must be changed. In order to make comparison, we still
write the specific heat at low-$T$ in the form
\begin{equation}
C_{V} = \mathcal{A}_{C_{V}}T^{2},
\end{equation}
where the coefficient
\begin{equation}
\mathcal{A}_{C_V}\!=\!N_b\mathcal{A}^{0b}_{C_V} +
N_f\mathcal{A}^{0f}_{C_V} + \mathcal{A}^{G}_{C_V},
\end{equation}
depends on temperature $T$. The susceptibility is linear in
temperature $T$ when the system is critical, but also deviates from
this behavior when the fermion becomes massive. It has the form
\begin{equation}
\chi_{b} = \mathcal{A}_{\chi_{b}}^{0b}T,
\end{equation}
for scalar boson $z$ and
\begin{equation}
\chi_{f} = \mathcal{A}_{\chi_{f}}^{0f}T,
\end{equation}
for fermion $\psi$. The susceptibility for Goldstone bosons
certainly vanishes since they do not couple to any magnetic field.

\subsection{$z$ scalar bosons}

When the system is not at the critical point, $r \neq 0$, all matter
fields carrying internal gauge charges are confined. Therefore,
their contributions to specific heat $C_{V}^{0b}$ and susceptibility
$\chi^{0b}$ strictly vanish at low-$T$. We only need to consider the
point of $r=0$, where the free energy is given by
\begin{equation}
f^{0b} = \frac{T}{2} \sum_{\omega_n}
\int\frac{d^{2}k}{4\pi^{2}}[\ln(k^2 + (\omega_{n}+H_{b}/2)^{2} +
m_{b}^{2}) + \ln(k^{2} + (\omega_{n}-H_{b}/2)^{2} + m_{b}^{2})] -
\frac{m_{b}^2}{g}.
\end{equation}
The parameter $m_{b}$ depends on $T$ and field $H_b$ as
\cite{Kaul}
\begin{eqnarray}
\!m_{b}\!=\!T\!\ln\!\left\{\frac{2\cos(H_b/2T)+1 +
\sqrt{(2\cos(H_b/2T)+1)^2-4}}{2} \!\right\}. \label{mh2} \nonumber
\end{eqnarray}
The contributions of $z$ bosons to specific heat and
susceptibility have been computed by Kaul and Sachdev \cite{Kaul},
with the results
\begin{eqnarray}
\label{eq:kaul_spheatb}
\mathcal{A}^{0b}_{C_V} &=& 6
\frac{8\zeta(3)}{10\pi} \approx 1.83661, \\
\label{eq:kaul_chib} \mathcal{A}_{\chi_{b}}^{0b} &=&
\frac{\sqrt{5}}{4\pi}\ln\frac{\sqrt{5}+1}{2} \approx 0.0856271.
\end{eqnarray}

\subsection{Massive fermions}

At low but finite temperature $T$, the fermion mass should be a
function $m(T,\mathbf{k})$ of $T$ and momenta $\mathbf{k}$. It is
usually difficult to get an analytical expression of
$m(T,\mathbf{k})$. As an approximation, we replace this function by
a constant $m$, which might be identified by the value $m(0,0)$ at
the limits $T=0, \mathbf{k}=0$. When $r = 0$, the free energy of
massive fermions is
\begin{eqnarray}
f^{0f} &=& - T \int \frac{d^2 k}{4 \pi^2} [ \ln (1 +
e^{-(\sqrt{k^2+m^2}+iH_f)/T})+\ln(1+e^{-(\sqrt{k^2+m^2}-iH_{f})/T})],
\end{eqnarray}
from which the specific heat of massive fermions is found to be
\begin{eqnarray}
\label{eq:spheat_f1} \mathcal{A}^{0f}_{C_V} &=&\frac{1}{\pi}\Big[-
(\frac{m}{T})^3\frac{e^{\frac{m}{T}}}{1+e^{\frac{m}{T}}}+3(\frac{m}{T})^2
\ln[1+e^{\frac{m}{T}}]+6(\frac{m}{T})\textrm{Li}_{2}[-e^{\frac{m}{T}}]
- 6\textrm{Li}_{3}[-e^{\frac{m}{T}}]\Big].
\end{eqnarray}
Here, $\textrm{Li}_{n}[z]$ is polylogarithm function.

At the limit $m \rightarrow 0$, it reduces to
\begin{eqnarray}
-\frac{6}{\pi}\textrm{Li}_{3}[-1] = 6\frac{3\zeta(3)}{4\pi}.
\end{eqnarray}

\begin{figure}[h]
  \centering
  \subfigure[]{
    \label{fig:spheat_f} %% label for first subfigure
    \includegraphics[width=3.0in]{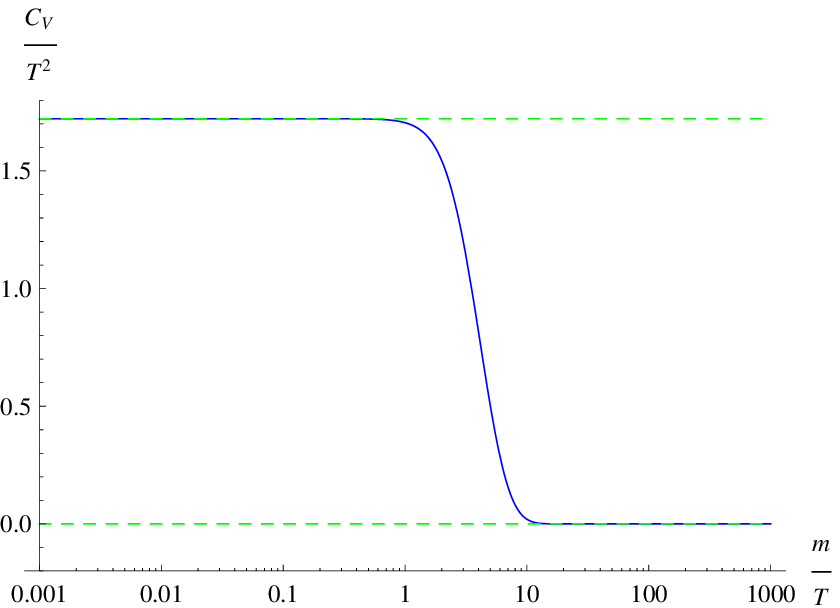}}
      \subfigure[]{
    \label{fig:chi_f} %% label for first subfigure
    \includegraphics[width=3.0in]{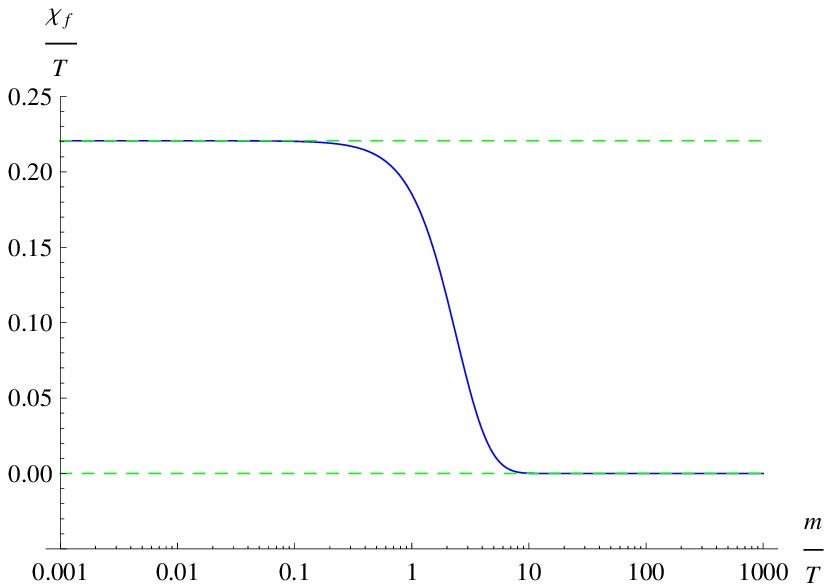}}
\caption{(a) Specific heat of massive fermions at very low $T$. The
top dashed line corresponds to $\mathcal{A}^{0f}_{C_V} =
\frac{9\zeta(3)}{2\pi}$. (b) Susceptibility of massive fermions at
very low $T$. The top dashed line corresponds to
$\mathcal{A}^{0f}_{\chi_f} = \frac{\ln2}{\pi}$.}
\label{fig:tq_m0} %% label for entire figure
\end{figure}

The dependence of $\mathcal{A}^{0f}_{C_V}$ on $\frac{m}{T}$ is depicted in
Fig. \ref{fig:spheat_f}. The susceptibility of massive fermions is
\begin{eqnarray}
\label{eq:chi_f} \mathcal{A}_{\chi_{f}}^{0f} &=&
\frac{1}{\pi}\big(-\frac{\frac{m}{T}e^\frac{m}{T}}{1 +
e^\frac{m}{T}} + \ln[1 + e^\frac{m}{T}]\big).
\end{eqnarray}
As $m \rightarrow 0$, it approaches $\frac{\ln2}{\pi}$. For finite
$m$, the dependence of $\mathcal{A}^{0f}_{\chi_{f}}$ on
$\frac{m}{T}$ is shown in Fig. \ref{fig:chi_f}. It is obvious that
the mass suppresses the fermion contributions to both specific heat
and susceptibility. Comparing with the contribution to specific heat
from the $z$ bosons, the contribution of massive fermions can be
simply neglected.

\subsection{Goldstone bosons}

Since the composite Goldstone bosons are neutral with respect to
gauge charges, they are not confined by the gauge potential and can
make finite contributions to certain observable quantities. At $r
\neq 0$, the Goldstone bosons are the only degrees of freedom at
very low-energy regime. Their contributions to specific heat are
nonzero and have the same expression at both $r=0$ and $r \neq 0$.
The free energy for Goldstone bosons is
\begin{eqnarray}
f^{G} = T\sum_{\omega_n}\int\frac{d^{2}k}{4\pi^{2}}[\ln(k^{2} +
\omega_{n}^{2})] = -\frac{2T^3\zeta(3)}{\pi}.
\end{eqnarray}
The specific heat is $\mathcal{A}^{G}_{C_V}T^{2}$ with
\begin{equation}
\label{eq:spheat_G} \mathcal{A}^{G}_{C_V} =
-\frac{1}{T}\frac{\partial^{2}f^{G}}{\partial T^{2}} =
12\frac{\zeta(3)}{\pi} \approx 4.591519.
\end{equation}
Their contribution to susceptibility is zero because they do not
couple to any magnetic field.

\subsection{Gauge contributions to physical quantities at critical point $r=0$}

At the deconfined critical point, $r=0$, the gauge field might give
rise to important contributions to specific heat and susceptibility.
The effective theory contains both temporal and spatial components
of the gauge field, so there is no singular corrections to the free
energy \cite{Kim97, Ran07}.

To calculate the gauge field corrections to free energy, we should
first obtain the polarization functions. The Feynman diagrams for
the fermion and scalar boson contributions to polarization functions
of gauge field are presented in Fig. \ref{fig:pi}. At finite
temperature, the one-loop vacuum polarization tensor are defined as
\begin{eqnarray}
\Pi_{\mu\nu}^{f}(p_{0},\mathbf{p}) &=& -\frac{N_{f}}{\beta}
\sum_{n}\int\frac{d^{2}\mathbf{k}}{(2\pi)^{2}}
\frac{\mathrm{Tr}[\gamma_{\mu}k\!\!\!/\gamma_{\nu}(p\!\!\!/+k\!\!\!/)]}
{k^2(p+k)^2}, \nonumber \\
\Pi_{\mu\nu}^{b}(p_{0},\mathbf{p}) &=& -\frac{N_{b}}{\beta}
\sum_{n}\int\frac{d^{2}\mathbf{k}}{(2\pi)^{2}}
\left[\frac{(p+2k)_{\mu}(p+2k)_{\nu}}{k^{2}(p+k)^{2}} -
\frac{\delta_{\mu\nu}}{k^{2}}\right].
\end{eqnarray}
Here, the Matsubara frequency is $p_{0} = 2m\pi/\beta$ for gauge
boson, $k_{0} = 2n\pi/\beta$ for scalar boson, and the frequency is
$k_{0} = (2n+1)\pi/\beta$ for Dirac fermion. Using the transverse
condition, $p^{\mu}\Pi_{\mu\nu}=0$, the polarization tensor can be
decomposed as
\begin{eqnarray}
\Pi_{\mu\nu}(p_{0},\mathbf{p}) &=&
[\Pi_{A}^{f}(p_{0},\mathbf{p})+\Pi_{A}^{b}(p_{0},\mathbf{p})]A_{\mu\nu}
+ [\Pi_{B}^{f}(p_{0},\mathbf{p})+\Pi_{B}^{b}(p_{0},\mathbf{p})]
B_{\mu\nu},
\end{eqnarray}
where
\begin{eqnarray}
A_{\mu\nu} &=& \left(\delta_{\mu 0} -
\frac{p_{\mu}p_{0}}{p^{2}}\right)\frac{p^{2}}{\mathbf{p}^{2}}
\left(\delta_{0\nu} - \frac{p_{0}p_{\nu}}{p^{2}}\right), \nonumber \\
B_{\mu\nu} &=& \delta_{\mu i}\left(\delta_{ij} -
\frac{p_{i}p_{j}}{\mathbf{p}^{2}}\right)\delta_{j\nu},
\end{eqnarray}
which satisfy the relationship
\begin{equation}
A_{\mu\nu} + B_{\mu\nu} = \delta_{\mu\nu} -
\frac{p_{\mu}p_{\nu}}{p^{2}}.
\end{equation}
Now the gauge boson propagator can be written as
\begin{eqnarray}
D_{\mu\nu}(p_{0},\mathbf{p}) &=&
\frac{A_{\mu\nu}}{p^{2}+\Pi_{A}^{f}(p_{0},\mathbf{p}) +
\Pi_{A}^{b}(p_{0},\mathbf{p})} +
\frac{B_{\mu\nu}}{p^{2}+\Pi_{B}^{f}(p_{0},\mathbf{p}) +
\Pi_{B}^{b}(p_{0},\mathbf{p})}.
\end{eqnarray}

The gauge contribution to free energy is given by:
\begin{eqnarray}\label{eq:f1a}
f^{1A} (\beta,H_{f}) = \frac{T}{2}\sum_{\epsilon_{m}}\int
\frac{d^{2}k}{(2\pi)^{2}}
\ln\Big[(p^{2}+\Pi_{A}^{f}+\Pi_{A}^{b})(p^{2}+\Pi_{B}^{f}+\Pi_{B}^{b})\Big].
\end{eqnarray}
The vacuum functions $\Pi_{A}^{f}$, $\Pi_{A}^{b}$, $\Pi_{B}^{f}$,
and $\Pi_{B}^{b}$ should be calculated explicitly. They are related
to the temporal and spatial components of vacuum polarization tensor
$\Pi_{\mu\nu}^{f}$ and $\Pi_{\mu\nu}^{b}$ by the identities
\begin{eqnarray}
\Pi_{A}^{f(b)} &=& \frac{p^{2}}{\mathbf{p}^{2}}\Pi_{00}^{f(b)}, \\
\Pi_{B}^{f(b)} &=& \Pi_{ii}^{f(b)} -
\frac{p_{0}^{2}}{\mathbf{p}^{2}}\Pi_{00}^{f(b)}.
\end{eqnarray}
In Ref. \cite{Kaul}, the temporal and spatial components of
polarization functions were obtained approximately. Indeed, these
functions can be computed exactly even in the presence of finite
magnetic field $H_{f,b}$ and fermion mass $m$. Using the methods
presented in Ref. \cite{Liwei}, we obtained the following
expressions for the fermion contribution to polarization functions
\begin{eqnarray}
\Pi_{00}^f(\epsilon_{m},\mathbf{k}) &=&
\frac{N_{f}}{\pi\beta}\int_{0}^{1}dx
\ln[4D_{m}(x,\epsilon_{m},\mathbf{k},H_{f})] -
\frac{N_{f}}{4\pi}\int_{0}^{1}dx (1-2x)
\epsilon_{m}\frac{\sin(x\beta \epsilon_m - \beta\theta
H_{f})}{D_{m}(x,\epsilon_{m},\mathbf{k},H_{f})} \nonumber \\
&& -\frac{N_{f}}{2\pi}\int_{0}^{1} dx
\frac{m^{2}+x(1-x)\epsilon_{m}^{2}}{E_{f}D_{m}(x,\epsilon_{m},\mathbf{k},H_{f})}\sinh(\beta
E_f), \nonumber
\end{eqnarray}
\begin{eqnarray}
&&\Pi_{11}^{f}(\epsilon_{m},\mathbf{k}) =
\Pi_{22}^{f}(\epsilon_{m},\mathbf{k})
\nonumber \\&=&
-\frac{N_{f}}{4\pi}\int_{0}^{1}dx \frac{(1-2x)
\epsilon_{m}\sin(x\beta \epsilon_m -\beta\theta
H_f)}{D_{m}(x,\epsilon_{m},\mathbf{k},H_{f})} -
\frac{N_{f}}{4\pi}\int_{0}^{1}dx \frac{x(1-x)(\mathbf{k}^{2} +
2\epsilon_{m}^{2})}{E_{f}D_{m}(x,\epsilon_{m},\mathbf{k},H_{f})}
\sinh(\beta E_{f}). \nonumber
\end{eqnarray}
The corresponding expressions for scalar bosons are
\begin{eqnarray}
\Pi_{00}^{b}(\epsilon_{m},\mathbf{k}) &=&
\frac{N_{b}}{2\pi\beta}\int_{0}^{1}dx \ln
\left[4J_{m}(x,\epsilon_{m},\mathbf{k},H_{b})\right] -
\frac{N_{b}}{4\pi}\int_{0}^{1}dx (1-x)\epsilon_{m}\frac{
\sin(x\beta\epsilon_{m}-\frac{\beta}{2}\theta
H_{b})}{J_{m}(x,\epsilon_{m},\mathbf{k},H_{b})} \nonumber \\
&& -\frac{N_{b}}{4\pi}\int_{0}^{1}dx
\frac{\frac{1}{2}x\mathbf{k}^{2} -
(x^{2}-\frac{3}{2}x+\frac{1}{4})\epsilon_{m}^{2}}{E_{b}
J_{m}(x,\epsilon_{m},\mathbf{k},H_{b})}\sinh(\beta E_{b}), \nonumber
\end{eqnarray}
\begin{eqnarray}
&&\Pi_{11}^{b}(\epsilon_{m},\mathbf{k}) =
\Pi_{22}^{b}(\epsilon_{m},\mathbf{k})
\nonumber \\&=&
-\frac{N_{b}}{4\pi}\int_{0}^{1}dx \frac{x\epsilon_{m}
\sin(x\beta\epsilon_{m}-\frac{\beta}{2}\theta
H_{b})}{J_{m}(x,\epsilon_{m},\mathbf{k},H_{b})} +
\frac{N_{b}}{8\pi}\int_{0}^{1}dx
\frac{x(1-x)(\mathbf{k}^{2}+\epsilon_{m}^{2})+\frac{1}{2}k_{1}^{2} -
x^{2}\epsilon_{m}^{2}}{E_{b}J_{m}(x,\epsilon_{m},\mathbf{k},H_{b})}
\sinh(\beta E_b). \nonumber
\end{eqnarray}
Here, we have defined a list of functions:
\begin{eqnarray}
D_{m}(x,\epsilon_{m},\mathbf{k},H_{f}) &=& \cosh^{2}(\frac{1}{2}
\beta E_{f}) - \sin^{2}(\frac{1}{2}x\beta\epsilon_{m} -
\frac{\beta}{2}\theta H_{f}), \nonumber \\
E_{f}(x,\epsilon_{m},\mathbf{k},m) &=&
\sqrt{m^{2}+x(1-x)(k^{2} + \epsilon_{m}^{2})}, \nonumber \\
J_{m}(x,\epsilon_{m},\mathbf{k},H_{b}) &=&
\cosh^{2}(\frac{1}{2}\beta E_{b}) -
\sin^{2}(\frac{1}{2}x\beta\epsilon_{m} - \frac{\beta}{4}\theta
H_{b}), \nonumber \\
E_b(x,\epsilon_{m},\mathbf{k}) &=& \sqrt{x(1-x)(k^2+\epsilon_m^2)}.
\nonumber
\end{eqnarray}
Note the scalar mass is taken to be $r=0$ since the gauge field is
deconfined only at this critical point.

The field $\lambda$ has dynamics only by integrating out the boson
fields $z$, which leads to the following free energy
\begin{eqnarray}\label{eq:f1l}
f^{1\lambda}(\beta,H_{b}) &=& \frac{1}{2\beta}
\sum_{\epsilon_{m}}\int \frac{d^{2}k}{(2\pi)^{2}} \ln\left[
\Pi_{\lambda}(\epsilon_{m},\mathbf{k})\right],
\end{eqnarray}
where the vacuum polarization, given by Fig. \ref{fig:pi_l}, at finite
temperature is
\begin{eqnarray}
\Pi_{\lambda}(\epsilon_{m},\mathbf{k}) =
\frac{N_{b}}{16\pi}\int_{0}^{1}dx \frac{1}{E_b} \frac{\sinh(\beta
E_b)}{J_{m}(x,\epsilon_{m},\mathbf{k},H_{b})}.
\end{eqnarray}

\begin{figure}[ht]
  \centering
  \subfigure[]{
    \label{fig:spheat_af} %% label for first subfigure
    \includegraphics[width=2.8in]{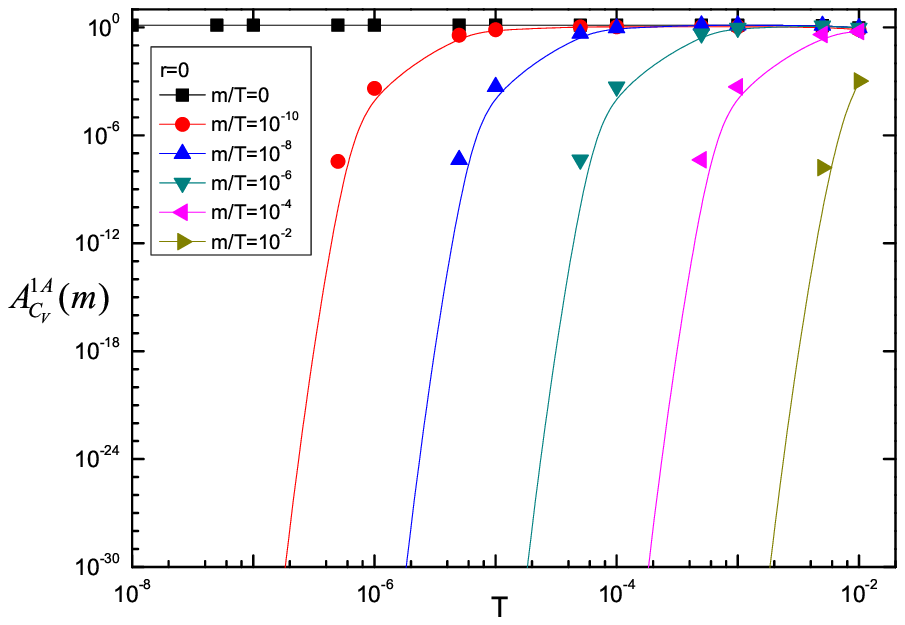}}
      \subfigure[]{
    \label{fig:chi_af} %% label for first subfigure
    \includegraphics[width=2.8in]{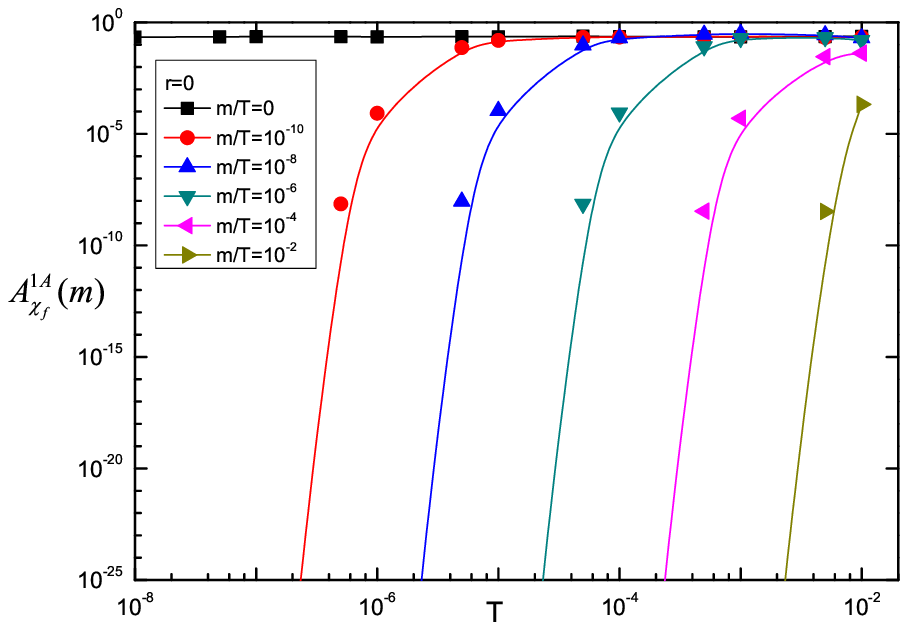}}
\caption{(a) The gauge field contribution to specific heat for
different fermion mass. (b) The gauge field contribution to
susceptibility for different fermion mass.}
\label{fig:tq_m4} %% label for entire figure
\end{figure}

\begin{figure}[ht]
  \centering
  \subfigure[]{
    \label{fig:spheat_lt} %% label for first subfigure
    \includegraphics[width=2.8in]{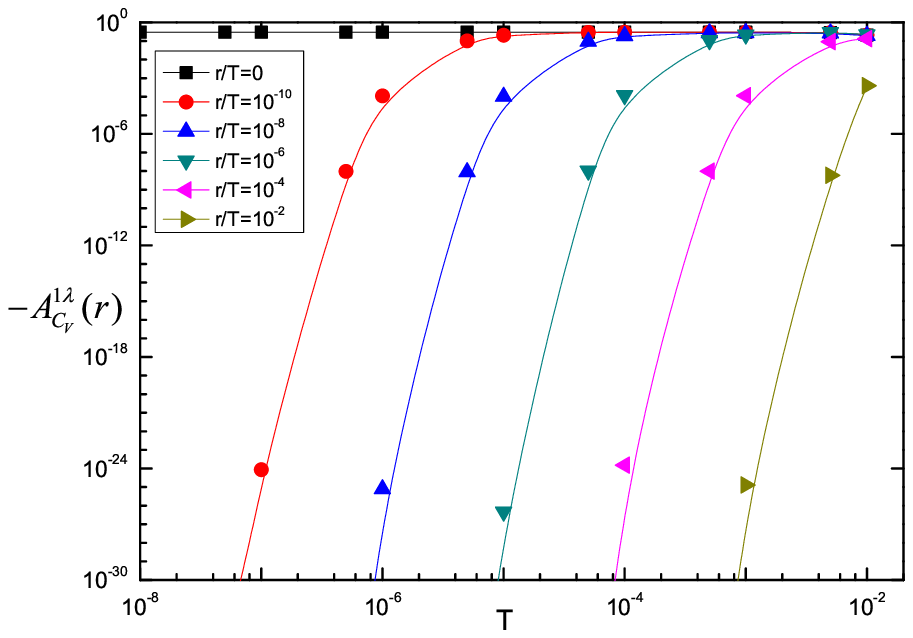}}
      \subfigure[]{
    \label{fig:chi_lt} %% label for first subfigure
    \includegraphics[width=2.8in]{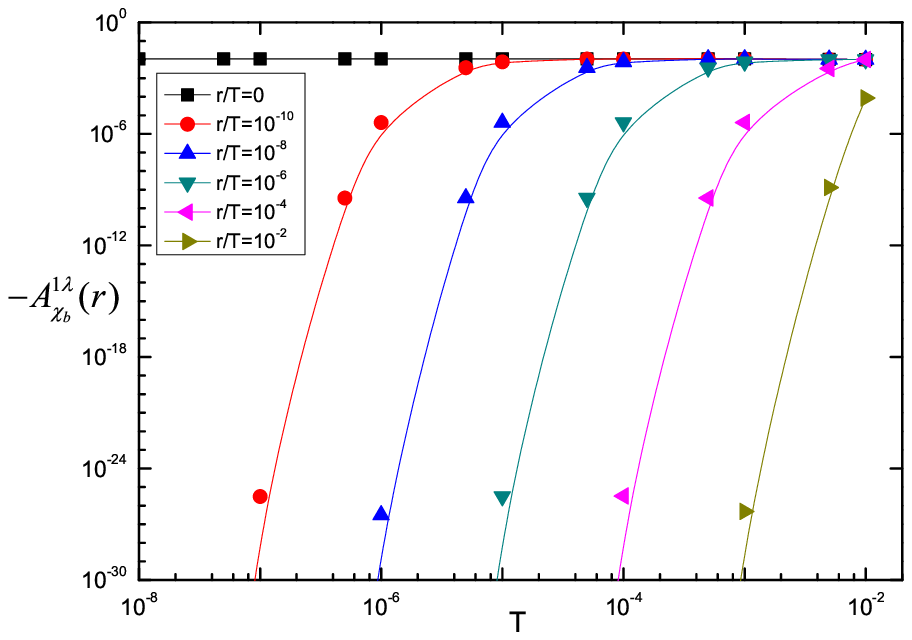}}
\caption{(a) The contribution of field $\lambda$ to specific heat
for different boson mass. (b) The contribution of field $\lambda$ to
susceptibility for different boson mass.}
\label{fig:tq_m5} %% label for entire figure
\end{figure}

Now we are ready to calculate the gauge field corrections to free
energy $f^{1A}$ and $f^{1\lambda}$ by inserting the polarization
functions $\Pi_{A}^{f}$, $\Pi_{A}^{b}$, $\Pi_{B}^{f}$, $\Pi_{B}^{b}$
into (\ref{eq:f1a}) and $\Pi_{\lambda}$ into (\ref{eq:f1l}). Then
the specific heat and susceptibility can be obtained by taking
derivatives of the free energy with respect to $T$ and $H_{f,b}$,
respectively. After numerical computations, the results for specific
heat and susceptibility are presented in Fig. \ref{fig:tq_m4} and
Fig. \ref{fig:tq_m5}. Apparently, the gauge contributions to
specific heat and susceptibility are significantly suppressed by the
mass of Dirac fermions. Here, in order to show the explicit results,
we use the absolute value of $|\mathcal{A}^{1\lambda}_{C_V}| =
-\mathcal{A}^{1\lambda}_{C_V}$ ($|\mathcal{A}^{1\lambda}_{\chi_b}| =
-\mathcal{A}^{1\lambda}_{\chi_b}$) instead of
$\mathcal{A}^{1\lambda}_{C_V}$ ($\mathcal{A}^{1\lambda}_{\chi_b}$).
Although $r=0$ at the critical point, for completeness we also
present the results for some magnitudes of finite $r$ in Fig.
\ref{fig:tq_m5}. If $N_{f} > N_{f}^{c}$, the fermions remain
massless and the matter fields carrying internal gauge charges are
not confined at finite $r$. In such state, the mass $r$ suppresses
the specific heat and susceptibility of $z$ bosons.

In summary, at the critical point $r=0$, both fermionc and gauge
field contributions to specific heat and susceptibility are
suppressed by the dynamical fermion mass. At low temperature, the
total specific heat receives contributions mainly from three parts:
free scalar boson $z$, the composite Goldstone boson, and the
fluctuation of field $\lambda$. Since the $z$ and Goldstone bosons
are both massless, these three contributions to specific heat are
all proportional to $T^{2}$, with coefficients $\mathcal{A}_{C_V}$
being temperature independent. However, the scalar bosons carry
nonzero internal gauge charge, while the Goldstone bosons are
neutral. Therefore, the former lead to a $\propto T$
susceptibility (to $H_{b}$), while the latter does not couple to any
magnetic field ($H_{f}$, $H_{b}$, or the physical $H$) and hence
make no contribution to susceptibility.

\section{Confinement and observable quantities at $r \neq 0$}
\label{sec:observable}

We now consider the Coulomb phase with $r \neq 0$. Once the
logarithmic gauge potential confines all particles that carry
internal gauge charges, the excitations appearing at low energy
regime must be neutral with respect to internal gauge charge. Now
the massive fermions can combine either with anti-fermions or with
massive scalar bosons carrying opposite gauge charges. In the former
case, the corresponding composite particles are just the massless
Goldstone bosons. In the later case, the separated spin and electric
charge degrees of freedom are re-combined by the confining internal
gauge potential. This implies that there are also electron-like
quasiparticles that carry both $1/2$-spin and $-e$ electric charge.
It is a nontrivial task to estimate their mass since this requires a
clear understanding about the mechanism of spin-charge
recombination. The issue of spin-charge recombination has been
investigated for many years by several authors \cite{Kim99, Rantner,
Mudry, Liu05}. Unfortunately, this issue is not at all clear. In the
absence of a reliable theory for spin-charge recombination, here we
assume a constant mass $m_{e}$. Further, the $z$ bosons can combine
with the $z^{\ast}$ bosons to form composite spin excitations
\cite{SenthilSci, Kaulprb, Kaul}. Both the electron-like fermionic
quasiparticles and the composite spin excitations are gapped, so
their contributions to specific heat and susceptibility (with
respect to real magnetic field $H$, rather than $H_{f,b}$) are
suppressed by their mass gaps $m_{e}$ and $\Delta_{\mathrm{spin}}$.
At very low temperatures, only the composite Goldstone bosons
contribute to specific heat (see equation (41)). However, although
having the same $T$-dependence at the point $r=0$, the magnitude of
specific heat in the Coulomb phase is much smaller than that at
$r=0$ since the contributions from massless $z$ bosons and the field
$\lambda$ all disappear at $r \neq 0$ due to confinement.

\section{Summary and discussion}
\label{sec:sum}

In summary, we have studied the possible mass generation of
initially massless Dirac fermions in a continuum U(1) gauge field
theory in (2+1) dimensions. It was found that a dynamical mass is
generated when the fermion flavor $N_{f}$ is less than $N_{f}^{c}$
which has quantitatively different values at the quantum critical
point $r=0$ and at $r \neq 0$. The gauge potential between charged
particles leads to logarithmic confinement when $r\neq 0$, but at
the critical point $r=0$ the matter fields carrying nonzero internal
gauge charge are deconfined. The low-energy properties of the system
at $r=0$ are quite different from those at $r \neq 0$. We discussed
some observable physical quantities, including specific heat and
susceptibility, of the system at low temperature and argued that
they can be used to judge whether the system stays at the quantum
critical point $r=0$ and whether the fermions acquire a dynamical
mass.

Besides specific heat and susceptibility, the happening of dynamical
fermion gap generation can also be readily seen from the transport
properties of the system. At very low temperature, the thermal
conductivity is mainly determined by the low-lying fermionic
excitations. If the Dirac fermions are massless, then they exhibit
universal thermal conductivity, $\frac{\kappa}{T} =
\frac{1}{3}\frac{v_{F}^{2} + v_{\Delta}^{2}}{v_{F}v_{\Delta}}$,
which is independent of impurity scattering rate $\Gamma_{0}$
\cite{Durst}. Once the Dirac fermions acquire a dynamical mass $m$,
then the thermal conductivity at zero temperature becomes
$\frac{\kappa}{T} = \frac{1}{3}\frac{v_{F}^{2} +
v_{\Delta}^{2}}{v_{F}v_{\Delta}}
\frac{\Gamma_{0}^{2}}{\Gamma_{0}^{2}+m^{2}}$, which is no longer
universal and is reduced by the fermion mass \cite{Gusynin}. If the
system is in the clean limit so that $\Gamma_{0} \ll m$, then the
thermal conductivity vanishes at very low temperature.

When chiral symmetry is dynamically broken (no matter $r=0$ or $r
\neq 0$), there are no gapless fermionic excitations at low
temperature. Therefore, in the limit $\Gamma_{0} \ll m$ the thermal
conductivity effectively vanishes at zero temperature and does not
have a universal linear term that would exist if the system contains
gapless Dirac fermions. The ground state of the system corresponds
to an interesting state of matter where the specific heat exhibits
behavior of gapless (bosonic) excitations but the thermal
conductivity has no residual linear term. In two recent works,
thermodynamic measurements carried out in
$\kappa$-(BEDT-TTF)$_{2}$Cu$_{2}$(CN)$_{3}$ found evidence for the
presence of gapless excitations \cite{Yamashita1}, while thermal
transport measurements carried out in the same material found good
evidence supporting the absence of gapless fermionic excitations
\cite{Yamashita2}. Although it is not known whether the ground state
of $\kappa$-(BEDT-TTF)$_{2}$Cu$_{2}$(CN)$_{3}$ can be described by a
continuum field theory analogous to the combined action
$\mathcal{S}_{f}+\mathcal{S}_{b}$, such seemingly contradicting
experimental facts are consistently realized by the chiral symmetry
breaking phase of action $\mathcal{S}_{f}+\mathcal{S}_{b}$ and can
be naturally understood by the mechanism discussed in the context.

The analysis presented above can also be applied to the Abelian
Higgs model, which may be considered as the relativistic version of
Ginzburg-Landau model, with Higgs field being the superconducting
order parameter. Although the $CP^{N_{b}-1}$ model is distinct in
physical contents from the Abelian Higgs model, its low-energy
property bears much resemblance to that of the latter.

\section*{Acknowledgments}

This work is supported by National Science Foundation of China No.
10674122.

\end{document}